\documentstyle[aps,prl,multicol,epsf]{revtex}
\begin{document}
\newcommand {\bfr}{\bf{r}}
\draft
\title{Vortex lattice melting in layered superconductors with 
periodic columnar pins}
\author{Chandan Dasgupta$^{1,}$ \cite{chandan} and 
Oriol T. Valls$^{2,}$ \cite{oriol}}
\address {$^{1}$Department of Physics, Indian 
Institute of Science, Bangalore 560012, India}
\address{$^{2}$Department of Physics
and Minnesota Supercomputer Institute,
University of Minnesota,
Minneapolis, Minnesota 55455-0149}
\date{\today}
\maketitle

\begin{abstract}

The melting transition of the vortex lattice in highly anisotropic,
layered superconductors with commensurate, periodic  columnar
pins is studied in a geometry where  magnetic field and  columnar
pins are normal to the layers. Thermodynamic properties and
equilibrium density distributions are obtained from numerical
minimizations of an appropriate free-energy functional. We find a line
of first-order transitions that ends at a critical point as the pin
concentration is increased. A simple Landau theory providing a
semi-quantitative explanation of the numerical results is proposed.
\end{abstract}

% insert suggested PACS numbers in braces on next line
\begin{multicols}{2}
\narrowtext
Thermodynamic and transport properties of type-II superconductors in
the mixed phase can be altered in a controlled manner by the
introduction of artificial pinning centers. Columnar
pinning in high-temperature superconductors (HTSCs), produced by
damage tracks from heavy-ion bombardment, has been extensively
studied experimentally\cite{exp1,exp2} and theoretically\cite{nv,rad}
because 
such pinning enhances the critical current.
A random array of such pins leads to a glassy phase,
known as Bose glass\cite{nv}. The interplay
between the lattice constant of the pin array and the intervortex
separation should produce\cite{df}
commensurability effects. Such effects have been observed in imaging
experiments\cite{harada} and magnetization measurements\cite{baert} 
on thin-film superconductors with periodic
arrays of artificially produced pinning centers.

In this Letter, we investigate the effects of a commensurate, periodic
array of columnar pins on the vortex-lattice melting
transition\cite{rmp} in highly anisotropic, layered, HTSCs in a
magnetic field. Both the field and the columnar pins are
assumed to be normal to the layers. We consider values of the 
field for which the pin areal density is smaller than that of the
vortex lines.  Although vortex lattice melting in HTSCs in the absence
of pinning has been studied extensively\cite{rmp}, not much is known
about the effects of columnar pins on this transition. Since columnar
defects produce strong pinning,  at
temperatures near the melting temperature of the vortex lattice in the
pure system each defect should pin a vortex. However, interstitial 
vortices, present when there are
more vortices than pins, should\cite{rad} undergo a melting transition
at a temperature slightly higher than the melting point of the pure
vortex lattice\cite{yyg}.  Evidence for such melting of interstitial
vortices has been found in experiments\cite{harada,baert} on thin-film
superconductors with periodic pinning, but the thermodynamics of this
transition has not been studied. 

Since the defect-pinned vortices
produce a periodic potential for the interstitial ones, the melting
transition of the latter provides a physical realization of
three-dimensional melting in the presence of an external periodic
potential. 
For small
concentrations of pinning centers, we find a first-order melting
transition from a crystalline solid to an inhomogeneous liquid. As the
pin concentration is increased, the transition temperature increases
and the latent heat and the jump in the crystalline order
parameter at the transition decrease. This line of first-order
transitions terminates at a critical point beyond which the
thermodynamic transition is replaced by a sharp crossover. This
critical point is a rare, experimentally realizable example of 
continuous melting in three dimensions. We show that
a simple Landau theory provides a semi-quantitative understanding of
most of our results.
Such melting transitions are of interest in other systems
such as atoms adsorbed on crystalline substrates\cite{nh}, colloidal
particles in interfering laser fields\cite{laser} and arrays of optical
traps\cite{traps}. Our work is of relevance to these systems
also.

Our study is based on a model free energy functional\cite{ry,gautam}
for a system of ``pancake'' vortices\cite{rmp} in a highly anisotropic
layered superconductor. The commensurate array of columnar pins is
accounted for 
by  an appropriate ``external potential'' term\cite{df} in the
free energy functional. By numerically minimizing a
discretized form of this  functional, we have studied the
effects of periodic pinning on the structure and thermodynamics
of the liquid and crystalline states of this system. 
We consider a layered superconductor with vanishingly small Josephson
coupling between layers (vortices on different layers are coupled via 
their electromagnetic interaction only). In this limit of 
effectively infinite anisotropy, which is
appropriate\cite{gautam} for
extremely anisotropic Bi- and Tl-based compounds, the energy 
of a system of pancake
vortices residing on the superconducting layers may be written as a sum
of anisotropic two-body interactions $v(n,r)$ where $n$ is the layer 
separation and $r$ is the separation in the plane of the layers. The
Fourier transform of  $v$ is\cite{gautam} 
\begin{eqnarray} 
\beta v({\bf k}) = \frac{2 \pi \Gamma
\lambda^2[k_{\perp}^2+(4/d^2)\sin^2(k_zd/2)]}{k_{\perp}^2[1+\lambda^2
k_{\perp}^2+4(\lambda^2/d^2) \sin^2(k_z d/2)]},
\label{inter}
\end{eqnarray}
with $\Gamma \equiv \beta d \Phi^2_0/8 \pi^2 \lambda^2\/$ 
and $\beta=1/k_BT\/$.
Here, $k_z (k_\perp)$ is the component of $\bf k$ perpendicular
(parallel) to the layer plane, $d$ is the layer spacing, $\lambda(T)$
the penetration depth in the layer plane, and $\Phi_0$ the flux
quantum. The intralayer potential $v(n=0,r)$ is repulsive and 
$\propto \ln(r)$, whereas the interlayer potential $v(n\ne 0,r)$, also
$\propto \ln(r)$, is attractive, weaker than the intralayer potential
by the factor $d/\lambda$, and decreases exponentially with $n$ as
$e^{-nd/\lambda}$. 
We use parameters appropriate to BSCCO i.e.  $\lambda(T=0) = 1500 \AA$
and $d = 15 \AA$, and assume a two-fluid $T$ dependence of
$\lambda$ with $T_c(0)=85$K.

In density functional theory\cite{ry,gautam,hm}, the free energy of a 
state in a density configuration specified
by $\rho(i,\bfr)$, the time averaged
areal density of vortices at point $\bfr$ on the $i$th
layer, is given in terms of equilibrium correlation functions of
the layered liquid of pancake vortices. We use the
Ramakrishnan-Yussouff free energy functional\cite{ry} which is
known\cite{gautam} to provide a quantitatively correct description of
the melting transition in our system in the 
absence of pinning. Since the potential produced by a set of straight
columnar pins perpendicular to the layers is the same on every layer,
the {\it time-averaged} density of vortices at any point $\bf r$ 
must be the same on all layers: $\rho(i,\bfr)= \rho(\bfr)$ for all
$i$. Then, the free energy per layer 
may be written in a
two-dimensional form:
\begin{eqnarray}
\beta (F[\rho] - F_0) &=& \int d^2r\left[\rho({\bfr})\ln 
\frac{\rho({\bfr})}{\rho_0} -
\delta\rho({\bfr})\right] \nonumber \\
&-&\frac{1}{2}\int d^2r \int d^2r^\prime \tilde{C}(|{\bfr} -
\bfr^\prime |)
\delta\rho({\bfr}) \delta\rho(\bfr^\prime) \nonumber \\
&+& \beta \int d^2r V_p({\bfr}) \delta\rho({\bfr}).
\label{ryfe}
\end{eqnarray}
Here, $\delta\rho({\bfr}) \equiv \rho({\bfr}) - \rho_0$, $F_0$ is the free
energy of the uniform liquid of areal density $\rho_0$ (= $B/\Phi_0$
where $B$ is the magnetic induction), $V_p({\bfr})$
is the pinning potential, and $\tilde{C}(r) \equiv \sum_n C(n,r)$,
where $C(n,r)$ is the {\it direct pair correlation function}\cite{hm} of a
layered liquid of pancake vortices. We use the results for $C(n,r)$ 
obtained \cite{gautam} from a hypernetted chain calculation\cite{hm}.

The pinning potential at point $\bfr$ is given by
$\beta V_p(\bf{r}) = {\sum_j} {\it V_0}(|\bf r - {\bf R}_j|)$,
where the sum is over all pinning centers located at the points $\{{\bf
R}_j\}$ on a plane, and $V_0(r)$, the potential at $\bfr$ due to a  
pinning center at the origin is assumed to have the truncated parabolic
form 
\begin{equation}
V_0(r) = -\alpha \Gamma (1 - r^2/r_0^2)  
\label{pin2}
\end{equation}
for $r \leq r_0$ and $V_0(r) = 0$ if $r >r_0$. Here, 
$r_0$ is the range 
and $\alpha$ is a ``strength''
parameter  chosen to ensure that on the average a pinning center
traps $\lesssim 1$ vortex in the temperature range of interest.

We find the minima of the free energy of Eq.(\ref{ryfe}) using a
methodology quite similar to that in our earlier
studies\cite{cd92,cdotv00} of hard-sphere systems. We discretize space 
by defining density variables $\{\rho_k\}$ at the sites of
a periodic grid, and use a gradient descent
method\cite{cd92} to locate the minima of the free energy of
Eq.(\ref{ryfe}) written as a function of $\{\rho_k\}$. The use of fast
Fourier transforms in the calculation of lattice sums\cite{pinsky}
speeds the computations and allows studies of larger systems.
To accommodate a triangular lattice, we use 
a triangular grid with periodic boundary conditions.

We have performed our studies for $B$ = 2kG and 3kG. We first checked
the results for crystallization without pinning. For this purpose, the
computational system was 
one triangular lattice unit cell  with lattice constant $a$ (all
lengths are in units of $a_0$, with $\pi a_0^2 \equiv 1/\rho_0=
\Phi_0/B$), and the spacing $h$ of the computational
grid was chosen to have the
values $a/n$ with $n$ = 16, 32, 64 and 128. 
The free energies of the crystal obtained\cite{dvfut} 
for all these values of $n$
are essentially the same, indicating that the effects of 
discretization
are minimal for $h \le a/16$. The equilibrium value $a_m$ of the lattice
parameter $a$ was determined by finding the value of $a$ that minimizes the
free energy at a given $B$ and $T$. 
The value of $a_m$ is found to be
slightly {\it higher} than the spacing of a perfect triangular lattice of
density $\rho_0$. This reflects the well-known 
result\cite{rmp} that the density of a vortex lattice {\it increases}
slightly at melting.

The free energy of the crystal crosses zero at the transition
temperature $T_c$ = 18.45K at $B$ = 2kG. As expected, $T_c$ is 
slightly higher than
that obtained from approximate treatments\cite{gautam} of the same
free energy. The entropy change $\Delta s$ 
per vortex is $0.29k_B$, and the jump in the order parameter $m$,
($m$ is the magnitude of the Fourier component of
$\rho(\bfr)$ at the shortest reciprocal lattice vector of the
triangular lattice) is $\Delta m=0.52$. The Lindemann parameter
calculated from the density distribution\cite{dvfut} at the crystalline
minimum 
is ${\mathcal L}$ = 0.26 at melting. Very similar results were obtained
for $B$ = 3kG ($T_c$ = 15.10K, $\Delta s$ = 0.28$k_B$, $\Delta m$ = 0.52,
${\mathcal L}$ = 0.25). The close agreement of these results with
those of other studies\cite{rmp,gautam} establishes the validity of our
numerical approach.

Next, we considered the effects of a single 
pin and a pair of pins on the liquid-state properties. The single-pin
calculation was done to fix the value of the parameter
$\alpha$ of Eq.(\ref{pin2}). Using
$r_0=0.1a_0$ ($\simeq 60 \AA$ for $B$ =2 kG), we find that in the
temperature range of interest ($T \approx$ 15-25K), 
the integrated density inside the range of the
pinning potential is close to one if $\alpha \approx 0.05-0.06$
($V_0(r=0) \approx$ 7-9).
%
%OTV, note that the Gamma defined here is 1/2 the gamma in our code
%
The density distribution near the center of the defect is
gaussian, $\rho(r) \propto \exp(-\alpha \Gamma r^2/r_0^2)$,
expected from Eq.(\ref{pin2}). Since pinning of multiple vortices at
radiation-induced defects is not found in experiments, higher values of
$\alpha$ 
%for which the integrated density exceeds one 
were not considered. Our two-pin calculations show\cite{dvfut} the
expected oscillatory behavior\cite{df} of the free energy as a function
of the distance between the pins.

To study melting in the presence of a commensurate, periodic array of
pins, we considered a triangular lattice of pins with spacing equal to
$la_m$ where $l$ is an integer. Thus the pin concentration is
$c \equiv 1/l^2$. The computational cell used was one unit cell of
the pin lattice (which contains $l^2$ unit cells of the vortex lattice)
with one pin located at the center of the first unit cell of the vortex
lattice. The value of $h$ was fixed at $a_m/64$. The crystalline (or liquid)
minimum of the free energy was located by starting the minimization from
a crystalline (or liquid-like) initial state, usually that obtained
at a nearby $T$. For both values of $B$ and smaller $c$,
($l$ = 10, 8, 7 and 6)
we found  a first-order
transition (two distinct local minima whose free energies cross at the
transition temperature). Results for $B$ = 2kG
are shown in Fig.\ref{fig2}. As expected, the presence of columnar pins
increases $T_c$. The discontinuities in $s$ and $m$ decrease as 
$c$ increases
(inset in Fig.\ref{fig2}) because
pinning-induced order in the liquid  
increases with $c$. The results for $B$ = 3kG are very similar, with
$T_c$ reduced by approximately 3.4K for all these values of $c$.

\vbox{
\vspace{-0.8cm}
\epsfxsize=8.0cm
\epsfysize=8.0cm
\epsffile{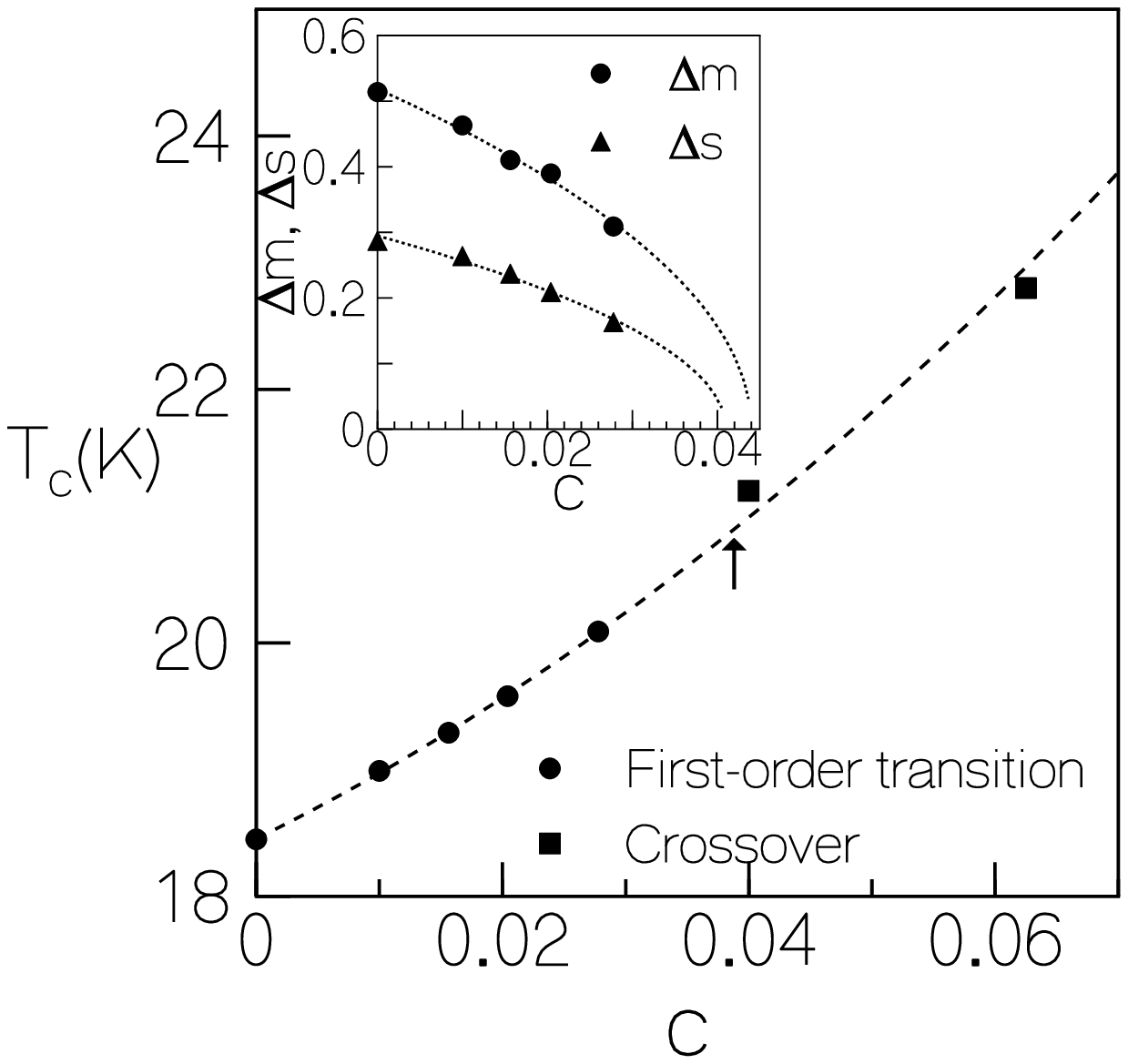}
\vspace{-0.6cm}
\begin{figure}
\caption{\label{fig2} The phase diagram ($T_c$
vs. pin concentration $c$) for $B$ = 2kG. The dashed line is a polynomial
fit. Inset: 
Entropy change $\Delta s$ (in units of $k_B$) and 
order parameter jump $\Delta m$ vs $c$. Dotted lines are fits to the
form $const. \times (c_c-c)^{1/2}$. The approximate position of the
critical point is indicated by the arrow.}
\end{figure}}

In Fig.\ref{fig3}, we have shown the variation of the local density
$\rho$ along a line joining two neighboring pinning centers for the
coexisting crystalline and liquid minima at the transition temperature for
$c=1/64$ and $B$ = 2kG. This plot brings out the differences between
the structures of the coexisting crystal and liquid phases and
illustrates the ability of our numerical method to provide detailed
information about the density distribution in inhomogeneous states.
In the liquid state, we find\cite{dvfut} the expected 
six-fold angular modulation\cite{df} of the density.

At larger values of $c$, ($l < $6)
the behavior found (for both values of $B$) is significantly
different. Here, the apparent minima obtained in ``heating'' runs
(increasing $T$ in steps from a crystalline state at low $T$ and using
the minimum obtained for the last $T$ as the input for the current
minimization) and in ``cooling'' runs (decreasing 
$T$ in steps from a liquid state at high $T$) have almost the same free
energy, but significantly different density
distributions\cite{dvfut} and values of $m$. At $c=1/25$, ($l$ = 5)
the difference in
the values of $m$ peaks at  $T=T_x\simeq$ 21.2K for $B$ = 2kG, 
as shown in Fig.\ref{fig4}. These results suggest
\vbox{
\vspace{-0.8cm}
\epsfxsize=8.0cm
\epsfysize=8.0cm
\epsffile{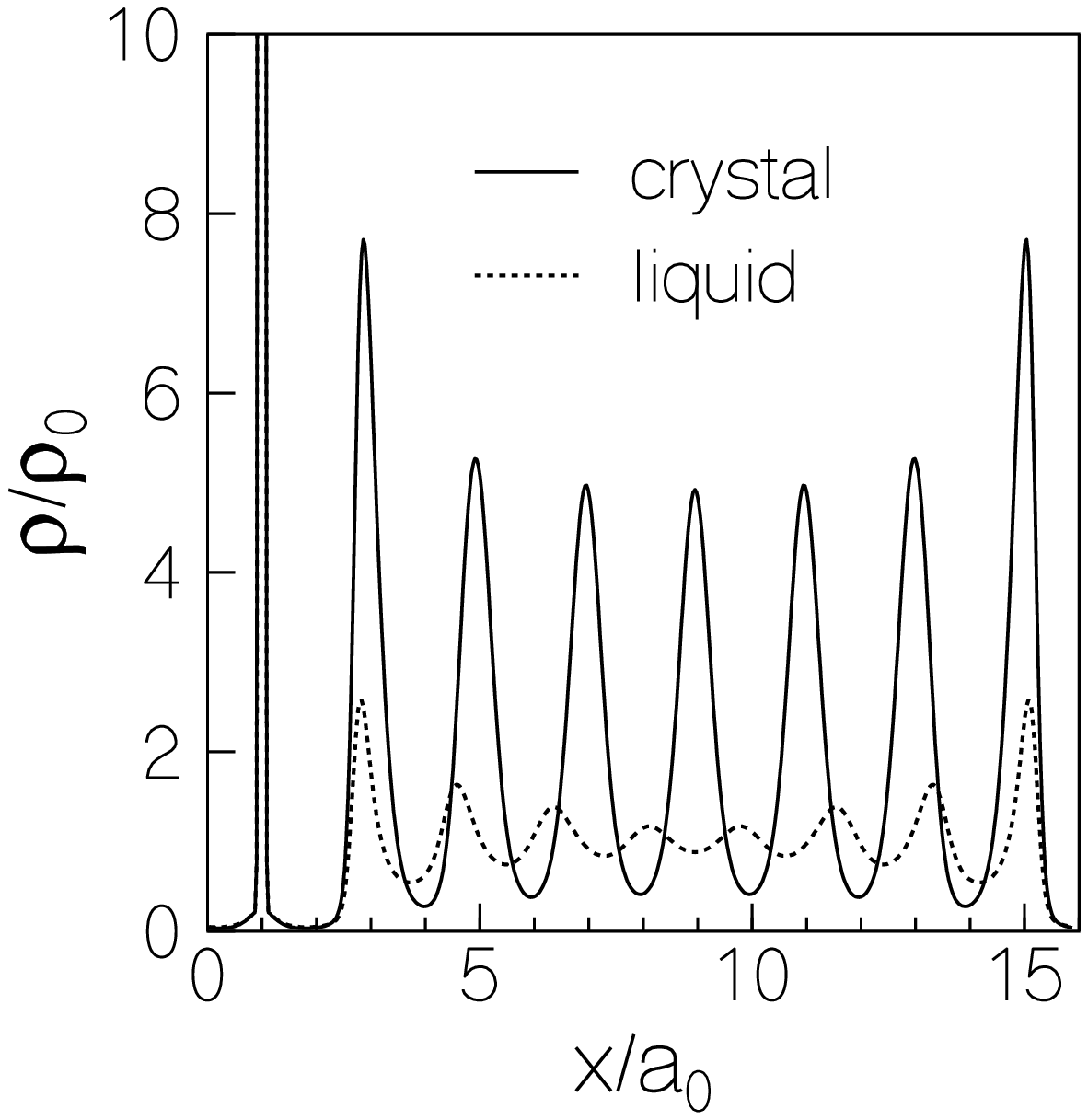}
\vspace{-0.6cm}
\begin{figure}
\caption{\label{fig3} The normalized local density $\rho/\rho_0$
along a line joining two adjacent pinning
centers. Data for the coexisting crystal (solid line) and liquid
(dotted line) for $c=1/64$, $B=2$kG are shown. The peak near $x/a_0=1$
is at the location of a pinning center. }
\end{figure}}
that $F$
has only one minimum that becomes very ``flat'' near $T=T_x$:
our minimization
routine, which assumes that a minimum has been reached when the gradient
of the free energy becomes smaller than a small convergence parameter, 
stops at slightly different points when approaching a very flat
minimum from different directions. To check this, we have calculated
the free energy of configurations $\{\rho_i(x)\}$ defined by  
$\rho_i(x) = x\rho_i^{(1)}+(1-x) \rho_i^{(2)}$,
where $\{\rho_i^{(1)}\}$ and $\{\rho_i^{(2)}\}$ are the configurations
obtained in heating (crystal) and cooling (liquid) runs,
respectively, and $0 \le x \le 1$ is a mixing parameter. A plot of the
free energy as a function of  $x$ (or of
$m(x)= xm^{(1)}+(1-x)m^{(2)}$, where $m^{(1)}$ and $m^{(2)}$ are the
order parameters in the two configurations) exhibits a minimum at
$x=x_0\sim 0.5$ at all temperatures. A typical plot, for $B$ = 2kG, $T$
= 21.2K, is shown in the inset of Fig.\ref{fig4}. In contrast, similar
plots for smaller values of $c$ exhibit a clear {\it maximum} at an
intermediate value of $x$. These results confirm that for $c=1/25$
or more, the
free energy has a unique minimum that lies between the configurations
obtained in heating and cooling runs. Thus, no first-order transition
occurs at $c=1/25$, and
the line of first-order transitions found for smaller
values of $c$ must end at a {\it critical point} lying between $c=1/36$ and
$c=1/25$. At $c=1/25$, the change from liquid-like to solid-like
behavior occurs as a sharp crossover. We identify the temperature
$T_x$, which coincides with the temperature at which the
temperature-derivative of the ``equilibrium'' value, $m(x_0)$, of the
order parameter peaks, as the crossover temperature. The sharpness of
the observed crossover (see Fig.\ref{fig4}) suggests that $c=1/25$,
$T=T_x\simeq$ 21.2K is close to the critical point for $B$ = 2kG, as
shown in Fig.\ref{fig2}. 
For $c=1/16$ the crossover is smoother. Our results for $B$ =
3kG are very similar, with $T_x \simeq$ 17.7K for $c=1/25$.

\vbox{
\vspace{-0.8cm}
\epsfxsize=8.0cm
\epsfysize=8.0cm
\epsffile{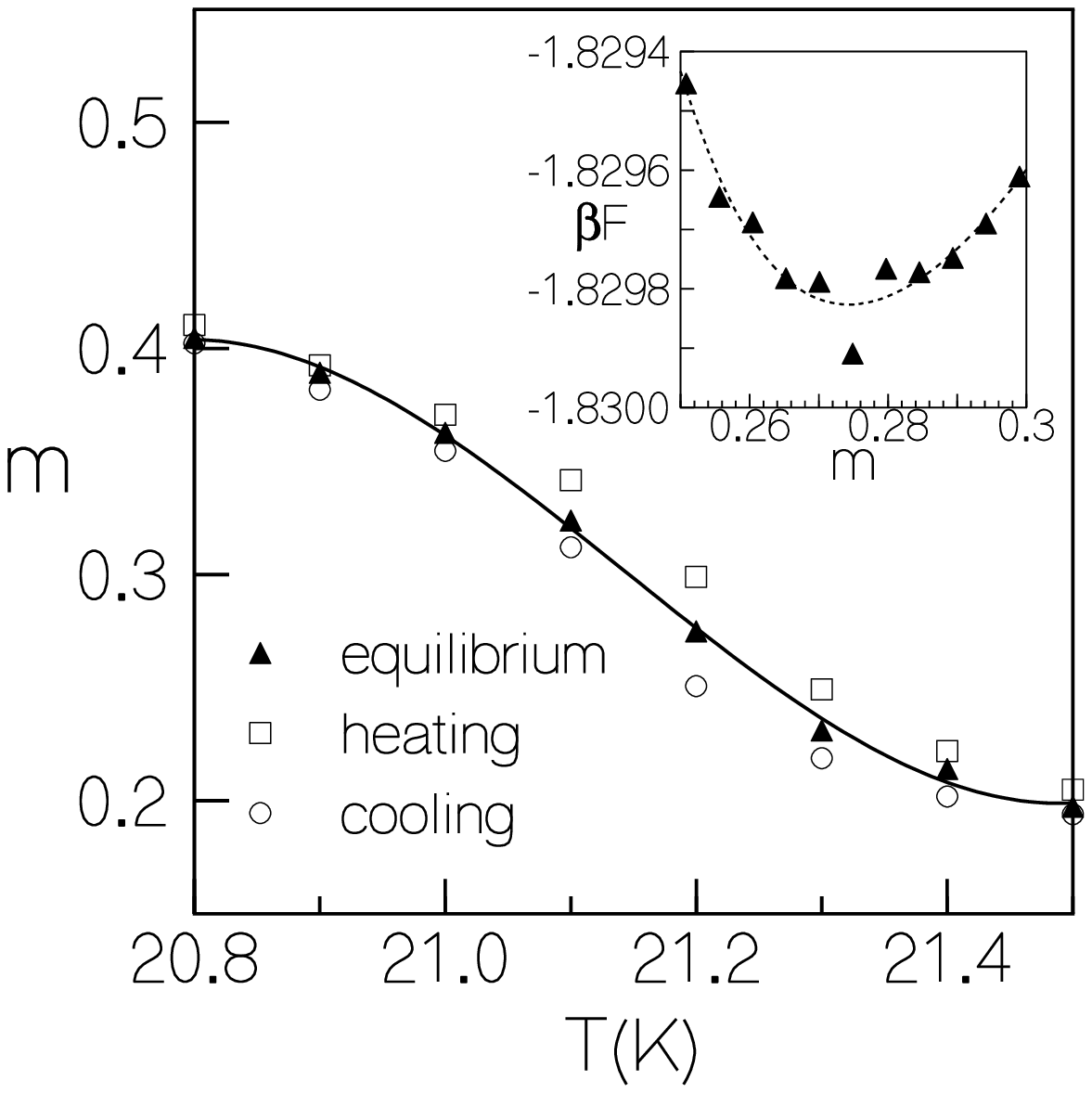}
\vspace{-0.6cm}
\begin{figure}
\caption{\label{fig4} The ``heating'', ``cooling'' and ``equilibrium''
values of the order parameter $m$ (see text)
as functions of  $T$
for $c=1/25$, $B$ = 2kG. The solid line is a polynomial fit to the
equilibrium data. Inset: Plot of the free energy vs. $m(x)$ at $T$ =
21.2K. The dotted line is the best fit to Eq.(\ref{landau}).}
\end{figure}}

We can describe the basic Physics underlying this phase diagram as follows.
In the presence of commensurate periodic pinning
the liquid and the crystal have the same 
symmetry.
Since the degree of order in the liquid increases with $c$, the liquid
and the crystal become indistinguishable beyond a critical value of
$c$. Thus,
it  is possible to go from one phase to the other without
crossing a phase boundary. A simple Landau theory 
illustrates this. Symmetry considerations\cite{frmc} suggest the
following Landau expansion for $F$:
\begin{equation}
\beta F = \frac{1}{2} a_2 m^2- \frac{1}{3} a_3 m^3 + \frac{1}{4} a_4
m^4 - hm, 
\label{landau}
\end{equation}
where $a_3$ and $a_4$ are positive constants, $a_2$ is a decreasing
function of temperature, and the ``ordering field'' $h$ is proportional to
the pin concentration $c$. This free energy
leads to a first-order transition for $h < h_c = a_3^3/27a_4^2$ and a 
critical point at $h=h_c, \, a_2=a_{2c}=a_3^2/3a_4$. The transition
temperature increases with $h$, and the latent heat and $\Delta m$
vanish as $(h_c-h)^{1/2}$ as $h$ approaches
$h_c$ from below. As shown in Fig.\ref{fig2} (inset), our data for
$\Delta s$ and $\Delta m$ are well-described by the form $\propto
(c_c-c)^{1/2}$ with $c_c$ close to 1/25. 
For a more quantitative comparison, we have fitted the $\beta F$ vs.
$m$ data for $B$ = 2kG, $T$ = 21.2K to the form of Eq.(\ref{landau})
(inset of Fig.\ref{fig4}).
The values of $a_{2c}$ and $h_c$, calculated from the best-fit values of
$a_3$ and $a_4$, are\cite{dvfut} less than 1.5\% lower than the 
best-fit values of $a_2$ and $h$,
indicating that the critical point for $B$ = 2kG is very close to 
$c=1/25$, $T$ = 21.2K. This explains the sharpness of the crossover
at $c=1/25$. Our results for $B$ = 3kG are very similar. 
The simple Landau theory, thus, provides a semi-quantitative account of
our density functional results and suggests that a critical point
lies very close to $c=1/25$ for both values of $B$. This 
critical point is analogous to the liquid-gas critical point in
mean-field theory. Fluctuations are expected to change this
correspondence because the symmetry of our order parameter is different
from that for the liquid-gas transition. 

Theoretical and experimental
investigations of the nature of this critical point
in highly anisotropic HTSCs would be welcome.
The critical point should be experimentally accessible: 
the  pin lattice spacing for
$B$ = 2kG, $c$ = 1/25 should be $\sim$ 0.55 $\mu$m, close to the
spacing of the radiation-induced pin array of Ref. \cite{harada}. 
Applications of our numerical method to other condensed matter systems
\cite{laser,traps} of a similar nature would also be very interesting.

\end{multicols}
\end{document}